# A hybrid optical-wireless network for decimetre-level terrestrial positioning

Jeroen C. J. Koelemeij[1], Han Dun[2], Cherif E. V. Diouf[2], Erik F. Dierikx[3], Gerard J. M. Janssen[2] & Christian C. J. M. Tiberius[2,✉]

[1]Vrije Universiteit Amsterdam, Amsterdam, The Netherlands. [2]Delft University of Technology, Delft, The Netherlands. [3]VSL, Delft, The Netherlands.

✉e-mail: C.C.J.M.Tiberius@tudelft.nl

**Global navigation satellite systems (GNSS) are widely used for navigation and time distribution[1,2,3], features indispensable for critical infrastructure such as mobile communication networks, as well as emerging technologies like automated driving and sustainable energy grids[3,4]. While GNSS can provide centimetre-level precision, GNSS receivers are prone to many-metre errors due to multipath propagation and obstructed view of the sky, which occur especially in urban areas where accurate positioning is needed most[1,5,6]. Moreover, the vulnerabilities of GNSS, combined with the lack of a back-up system, pose a severe risk to GNSS-dependent technologies[7]. Here, we demonstrate a terrestrial positioning system which is independent of GNSS and offers superior performance through a constellation of radio transmitters, connected and time-synchronised at the sub-nanosecond level through a fibre-optic Ethernet network[8]. Employing optical and wireless transmission schemes similar to those encountered in mobile communication networks, and exploiting spectrally efficient virtual wideband signals, the detrimental effects of multipath propagation are mitigated[9], thus enabling robust decimetre-level positioning and sub-nanosecond timing in a multipath-prone outdoor environment. This work provides a glimpse of a future in which telecommunication networks provide not only connectivity, but**

**also GNSS-independent timing and positioning services with unprecedented accuracy and reliability.**

The limitations of GNSS in urban areas are difficult to overcome. The limited radio bandwidth (~20 MHz) of GNSS systems implies metre level positioning errors due to multipath propagation, i.e. signals reflected off buildings and objects that interfere with the direct line-of-sight (LoS) signal[10], while an obstructed view of the sky will impair even advanced GNSS receivers aided by inertial navigation systems[5,6]. These restrictions not only limit navigation applications: GNSS receivers are also widely used as time references[3], enabling synchronisation of local clocks to coordinated universal time (UTC) with residual time errors of several nanoseconds at best. However, emerging technologies, such as quantum communication, require sub-nanosecond time synchronisation of (quantum) network nodes[11]. Moreover, even if GNSS errors could be reduced by the required 1-2 orders of magnitude, yet another bottleneck exists in the form of the unencrypted and relatively weak GNSS radio signals, which may be jammed, spoofed, or even forged to convey false time and position information. Amidst mounting concerns about the vulnerability of GNSS-dependent systems to (un)intentional disruptions[7], network operators[3] and governments[12,13,14] have begun looking into possible back-up systems for GNSS.

**Alternative timing and positioning**

Several systems have been developed to cope with the limitations of GNSS. These include low Earth orbit satellite constellations[14,15] and local terrestrial positioning systems[14] based on GNSS-inspired signal structures with augmentations [one example being the Locata system, which employs two code-division multiple access (CDMA) signals with a 10 MHz chip rate on two frequencies, 51 MHz apart, within the 2.4 GHz band[16,17]] for improved signal power, coverage, and resilience (if combined with GNSS). Other systems make use of two-way ranging to Wi-Fi access points for positioning at the metre level[14], or involve local networks of ultra-wideband radio transceivers to reduce positioning and timing errors due to multipath to the decimetre level[18] and sub-nanosecond regime[20], respectively. However, these systems rely on two-way communication or radio transmissions from the mobile transceiver to a sensor infrastructure. Users thus have to reveal their presence and position, which is less privacy-friendly than the use of downlink-only (silent) GNSS receivers. Downlink-only positioning methods, like GNSS, also scale more favourably than two-way methods, for which the network overhead increases with the number of users. While each of the abovementioned solutions overcomes some of the GNSS limitations – sometimes leading to spectacular improvements – they require their own application-specific infrastructure, which forms a major obstacle to large-scale deployment. Established wireless-network-based (and therefore scalable) positioning methods, on the other hand, are still outperformed by GNSS[14,21].

Here, we describe a terrestrial networked positioning system (TNPS) based on a hybrid optical-wireless telecommunication infrastructure, which provides positioning in urban environments with centimetre to decimetre level uncertainty as well as sub-nanosecond time synchronisation. The system operates independently of GNSS, and many of the risks associated with GNSS are eliminated by the use of a fibre-optic infrastructure in combination with a constellation of terrestrial transmitters that enable received signal power levels one million times higher than typically received GNSS-signal powers. The TNPS demonstrated here is depicted schematically in Fig. 1 and Extended Data Fig. 1. A TNPS testbed was erected at an urban outdoor test site, The Green Village (TGV), located at the campus of Delft University of Technology (Fig. 1a, Extended Data Fig. 2). Six radio transmitters (Tx-1 … Tx-6), dispersed over an area of 660 m$^2$, play the role of anchor stations similar to the satellites of a GNSS constellation. Like in GNSS, the key concept that underpins the TNPS is the accurate measurement (by a mobile receiver) of arrival times of the various radio signals transmitted by the constellation. Time-of-arrival measurements are converted to ranges, and subsequently used to locate the receiver relative to the constellation through multilateration. Such measurements necessitate tight synchronisation of the internal clocks of the transmitters:

with radio signals propagating at nearly the speed of light in a vacuum, $c$=299,792,458 m/s, a time error of 1 ns translates to a 3-dm range error. We achieve sub-nanosecond synchronisation of the transmitters by use of White Rabbit (WR), a timing protocol that is implemented in optical 1.25 Gb/s Ethernet (8-bit/10-bit encoding) equipment[8]. Thus, the internal clocks of all WR network nodes are synchronised to a common reference clock so that each node replicates the time and frequency of the reference clock. Such synchronisation is achieved irrespective of fibre-optic and electrical signal propagation delays, which are quasi-continuously estimated by the WR protocol from measured round-trip delays, and electronically compensated. As shown in Figs. 1b and 2a, each transmitter in the TNPS network receives timing signals [in the form of 10 MHz and 1 pulse-per-second (PPS) electrical signals] from its associated WR timing node (TN), so that joint radio transmissions can be timed with an uncertainty of about 0.2 ns (see Methods for details). The reference clock for the WR network of our testbed is located at the Dutch metrology institute, VSL, and is connected to the TNPS testbed through approximately 2.5 km of installed telecommunication optical fibre. At VSL, a WR device is configured as WR Grandmaster (GM), taking its reference time and frequency directly from the timescale UTC(VSL), the Dutch realisation of Coordinated Universal Time. Through the WR network, the clocks of all nodes of the TNPS network are phase-coherently linked to UTC(VSL) and traceable to the base unit of time in the International System of Units (SI), the second.

**Radio signal processing and positioning**

The TNPS transmitters and receiver (Rx) are based on programmable universal software radio peripherals (USRPs). Tx stations transmit bursts of modulated radio signals for positioning, navigation and timing (PNT) at a carrier frequency of 3.96 GHz and an effective radiofrequency (RF) bandwidth (BW) of 160 MHz (Methods). The large BW helps resolve LoS signals from signals that are reflected by the various nearby buildings and objects at the test site. The signal structure and relative timing of the transmissions by Tx-1 ... Tx-6 are depicted in Figs. 2a,b. PNT signals are encoded using orthogonal frequency division multiplexing (OFDM); see Fig. 2b. To this end we divide the 160 MHz signal BW into $M$=16 subbands of 10 MHz, each containing $N$=64 subcarriers, thus creating a multiband-OFDM signal. Good positioning performance can also be achieved using sparse signal bands (in our case employing only 7 out of 16 subbands), provided that the bands at the bandwidth edges are included[22]. Such sparse bands resemble the non-contiguous bands or combinations of separated bands that are frequently encountered in mobile communication systems.

The position of the receiver is found from the arrival times of the OFDM signals as follows. The first step consists in determining the delay, $\hat{\tau}_r^i$, between transmission by Tx-$i$ and reception by the receiver (Methods). From $\hat{\tau}_r^i$ the pseudo-range, $\rho_r^i$, is derived, which (at a given time $t$) can be expressed as

$$\rho_r^i(t) = c\hat{\tau}_r^i(t) = d_r^i(t) + c\tau_h + c\tau_h'^i + c\delta_r(t) + \varepsilon_r^i(t). \tag{1}$$

In Eq. (1), $d_r^i(t)$= |$\mathbf{x}_r(t) - \mathbf{x}^i$| stands for the time-dependent geometric propagation distance between Tx-$i$ and Rx (with $\mathbf{x}_r(t)$ and $\mathbf{x}^i$ the three-dimensional position vectors of the receiver and Tx-$i$, respectively), $\tau_h$ denotes the common hardware delay of all Tx-Rx links (which arises because all transmitters are based on identical USRPs, antennas, and cables), and $\tau_h'^i$ are small residual Tx-$i$-specific hardware delays; both $\tau_h$ and $\tau_h'^i$ were determined separately (Methods). Furthermore, $\delta_r(t)$ represents the unknown time-dependent clock offset of the receiver, and $\varepsilon_r^i(t)$ is a term that lumps together all errors due to multipath, unknown residual hardware delays, and timing jitter. From the received OFDM signals the receiver determines for each transmitter the delay $\hat{\tau}_r^i$ and pseudo-range $\rho_r^i$, from which the receiver position $\mathbf{x}_r(t)$ and clock offset $\delta_r(t)$ are subsequently solved by maximum likelihood estimation[22,23] (MLE) (Methods). In addition to the OFDM time-delay (TD) positioning described above, we performed positioning with enhanced precision based on carrier-phase (CP) measurements (Methods).

**Positioning and timing performance**

To assess the TNPS performance in an urban environment, static and slow kinematic tests (with the receiver unit carried by either a trolley or a car) were conducted at TGV, and seven test runs were performed in total (see also Methods and Extended Data Fig. 3). An optical land-surveying ground-truth (GT) system was deployed which provides position reference measurements with an estimated root-mean-square error (RMSE) of about 2.5 cm (Methods). Figure 3a shows OFDM-TD and GT position solutions during one run where the receiver was moved on a trolley along a 30 m long trajectory. Figure 3b displays the position errors for this run (as well as their mean and standard deviation), indicating good agreement with the GT results, and an overall empirical RMSE of 10.2 cm in the east-west (EW) direction, and 7.4 cm in the north-south (NS) direction (Table 1). Results from full-band and sparse-band OFDM-TD measurements are compared with CP results in Fig. 3c. We find that sparse-band and full-band OFDM-TD positioning have similar decimetre-level performance, as expected from the similar values for the theoretical Cramér-Rao lower bound (CRLB) for both methods[9,22]. CP positioning, however, delivers centimetre-range positioning: using ambiguity-float and ambiguity-fixed solutions (Methods), we achieved RMSEs as low as 4-5 cm and 2-4 cm (EW-NS directions), respectively (Table 1).

The spread in the record of position errors is partly due to jitter and wander in the transmission times of the PNT signals. We experimentally determined an upper limit of 0.2 ns for these timing variations, which consequently contribute at most 3-5 cm (EW-NS) to the RMSE. The empirical RMSE of the TD positions, however, is appreciably greater, which we attribute to multipath effects. Indications of multipath are also found in the measured pseudo-ranges; Fig. 4a displays residual range errors (defined as TD range minus GT range) for each Tx-$i$. Apparently, range errors of several decimetres for a single transmitter occur. However, owing to the spatial diversity and redundancy of the constellation of transmitters, the weight of such errors in the position solution is reduced so that an RMSE below 10 cm is achieved.

Multipath is therefore a limiting factor in our TNPS, albeit at a much lower level – decimetre instead of metre – than in GNSS positioning by virtue of the much wider BW employed. This is illustrated in Fig. 4b, where theoretical multipath error envelopes for TD-ranging using the Global Positioning System (GPS) (L5 signal, 20.46 MHz BW) and the TNPS are shown. The maximum multipath error visible in Fig. 4b is 4 m for GPS, compared to 0.5 m achieved through the TNPS using full-band (160 MHz BW) as well as sparse-band OFDM (the latter occupying only 70 MHz of BW). CP ranging can be even less sensitive to multipath, because the sum phasor of multipath components with a combined amplitude smaller than that of the LoS component will lead to a phase error of at most $\pi/2$, corresponding to a range error of less than 1.89 cm at 3.96 GHz.

Apart from multipath, radio navigation systems suffer from scenarios where the direct LoS to the transmitter is blocked, but where non-LoS (NLoS) multipath signals are still detected by the receiver, thus leading to position errors. Various strategies to detect and mitigate NLoS propagation in (ultra)wideband systems exist[24,25]. A conservative measure against NLoS is to omit the pseudo-range of an impaired base station from the positioning algorithm. To illustrate the robustness of our TNPS to such NLoS-implied omission of one or two base stations, we have computed position solutions for one of the runs at TGV for various reduced constellation sizes; see Extended Data Fig. 4 and Extended Data Table 1. Owing to the redundancy of the constellation, decimetre-level positioning remains possible in most cases if one or two arbitrary base stations are removed. One exception is base station Tx-6, whose exclusion renders a large section of the test run outside the remaining constellation's coverage area, where – due to a mechanism known as geometric dilution of precision – positioning errors of several decimetres to up to a few metres may readily occur. Errors of a few decimetres may also occur for situations in which three base stations – half of the constellation – are excluded. For CP positioning, at least five base stations are required[22], so that only one base station may be lost from our constellation. Our experimental data show that CP positioning with centimetre-range uncertainty remains possible also if one base station is removed (Extended Data Table 2).

The TNPS also enables accurate wireless time distribution, because the Rx clock offset with respect to WR network time is determined with sub-nanosecond uncertainty (Table 1 and Methods). The uncertainty is limited by errors due to multipath, the limited precision with which hardware delays were determined, and small residual misalignments of the time bases of the Tx stations. In our TNPS, WR network time is traceable to UTC(VSL), and the traceability was independently verified using a portable atomic clock with a standard uncertainty of 50 ns, limited by the drift of the portable clock (Methods). However, WR supports significantly smaller time uncertainties. This is illustrated by the time stability (TDEV[26]) of the roundtrip comparison between WR-SL and UTC(VSL), which was typically below 10 ps for averaging times ranging from 2 minutes to up to 115 days (Fig. 4c). The TDEV shown in Fig. 4c corresponds to a modified Allan deviation (MDEV[26]) of about $1\times10^{-18}$ after $10^7$ s of integration (Extended Data Fig. 5). The roundtrip time offset was 6 ns, which we attribute to uncalibrated delay asymmetries in the WR hardware and chromatic dispersion of the optical fibre. Well-developed methods[27,28,29] exist to calibrate such delay asymmetries, making it is possible to synchronise WR devices to within a fraction of a nanosecond even over distances in excess of 100 km[30,32]. Therefore, our TNPS might be used to broadcast national realisations of UTC over the air with sub-nanosecond uncertainty, an improvement over common GNSS time distribution techniques by two orders of magnitude[33].

**Compatibility with existing networks**

The TNPS reported here represents a milestone towards robust, network-based decimetre-level positioning and wireless sub-nanosecond time distribution in urban environments, based on signals that are significantly more difficult to tamper with than GNSS radio signals (because the relatively high signal levels and anticipated density of base stations will limit jamming and in particular spoofing attacks to a small geographic area). However, the potential of any PNT technology to serve as a GNSS backup or next-generation positioning system also depends on the effort needed for its deployment. We here point out several crucial features of our TNPS which enhance its compatibility with existing telecommunication networks, thus reducing the barrier to implementation. First, there is the high technological readiness level of WR: the WR protocol has been standardised[34], WR equipment has been commercially available for over one decade and, owing to its relatively modest bitrate, WR can operate (without loss of performance) at wavelengths in the fibre-optic spectrum that experience relatively large attenuation. Such wavelengths are often available as they are less suitable for high-capacity data transmission systems. Employing wavelength division multiplexing (WDM), WR operating in such unused wavelength bands has already been successfully integrated into long-distance fibre-optic networks that simultaneously carry live high-capacity data traffic, showing no interaction between the WR signals and the other data traffic, and achieving sub-nanosecond timing performance[30,31,32].

Second, our TNPS system may be implemented in ways that circumvent the scarcity of radio bandwidth. Present terrestrial PNT systems primarily make use of unlicensed frequency bands (e.g. the Industrial, Scientific and Medical band at 2.4 GHz in which Wi-Fi operates), while some other systems operate in proprietary spectrum (such as the NextNav Metropolitian Beacon System, which operates in an 8-MHz band around 923.75 MHz[14]). The sparse-band radio system demonstrated here could be designed such that its spectral footprint matches the spectra licensed to mobile network operators. These consist typically of several bands of 5-15 MHz BW distributed over the frequency range 600-3500 MHz, so that a sparse ultra-wide virtual BW may be achieved[22]. In addition, the OFDM modulation format is similar to that used in 4G and 5G radio access networks as well as Wi-Fi networks, suggesting that the OFDM PNT signals may be transmitted by present-day mobile base stations (in line with current and evolving standards such as 3GPP 36.305 for 4G and 38.305 for 5G) and Wi-Fi access points (assuming these are synchronised using WR or similar protocols through the optical network, or over the air through two-way OFDM signal exchange[35] with nearby TNPS reference transmitters). In our prototype TNPS, at a PNT message rate of 1 kHz, up to 94% of the telecommunication capacity remains available; however, this figure may improve to well above 99% by straightforward modifications of the PNT signal structure (such as joint PNT transmissions employing CDMA instead of TDM, or a reduced PNT message rate), and without loss of positioning performance (Methods). With the continuing densification of cellular networks and wireless access points, the density of such integrated telecom/TNPS transmitters may ultimately become sufficient to achieve city-wide or even nation-wide coverage, with a high level of robustness against MP and NLoS signals.

**Acknowledgements** This research is funded through the Dutch Research Council (NWO) under grants 12346 and 13970, with additional support from KPN, VSL, OPNT, and Fugro. We acknowledge support on the optical infrastructure from L. Boonstra, T. Theijn, and R. Smets, from L. Colussi and F. van Osselen on obtaining the 3.96 GHz experimental license, and R. Tamboer and T. Jonathan on realizing the testbed at TGV.

**Author Contributions** Conceptualisation, J.C.J.K., G.J.M.J., C.C.J.M.T.; methodology, J.C.J.K., H.D., C.E.V.D., E.F.D., G.J.M.J., C.C.J.MT.; prototype system development, H.D., C.E.V.D.; prototype deployment and field trial (experiment), H.D, C.E.V.D., E.F.D., G.J.M.J., C.C.J.M.T.; measurement data processing, analysis and validation, H.D., C.E.V.D.; writing–original draft preparation, J.C.J.K.; writing–review and editing, J.C.J.K., H.D., C.E.V.D., E.F.D., G.J.M.J., C.C.J.M.T.; visualisation, J.C.J.K., H.D., C.E.V.D.; project administration and funding acquisition, J.C.J.K., G.J.M.J., C.C.J.M.T. All authors have read and agreed to the published version of the manuscript.

**Competing interests** The authors declare the following competing interests: J.C.J.K. is co-founder and shareholder of OPNT bv. The authors declare no further competing interests.

**Correspondence and requests for materials** should be addressed to C.C.J.M.T.

**Reprints and permissions information** is available at http://www.nature.com/reprints.

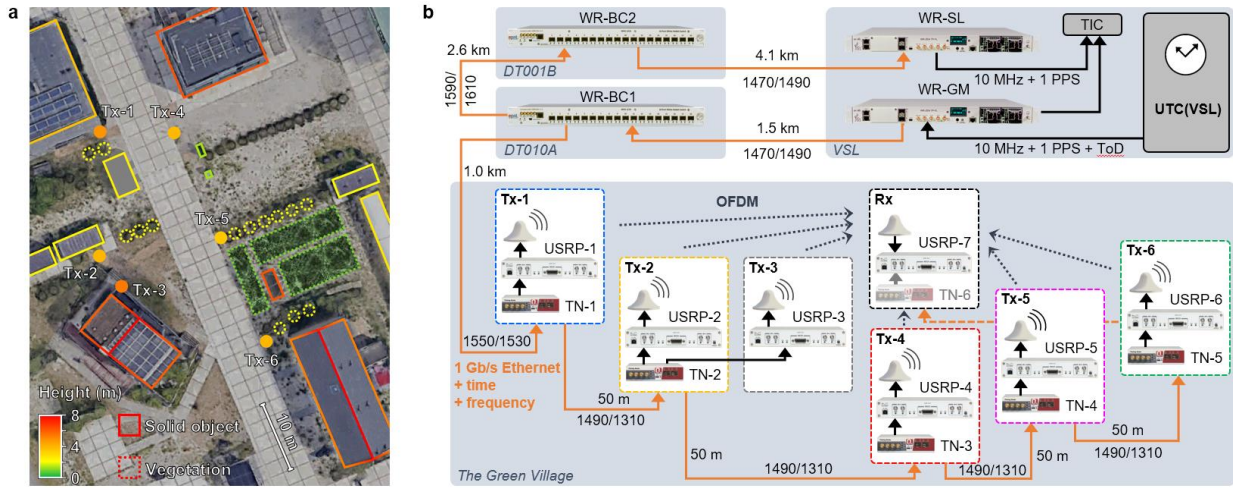

**Fig. 1| TNPS testbed. a**, Aerial view (Google Earth) and heightmap of The Green Village (TGV) test site, showing the Tx locations and several buildings and objects. **b**, Schematic layout of the TNPS testbed. A WR Grandmaster (GM) is synchronised to UTC(VSL) by electrical signals (black arrows), and transfers time-of-day (ToD) and frequency to WR boundary clocks (BCs), located in data centres DT010A and DT001B, and WR timing nodes (TNs) at TGV through a local telecommunication optical fibre network (orange arrows, fibre lengths are indicated). WR communication employs downstream/upstream wavelengths as indicated (in nanometres); wavelength channels are multiplexed into the optical fibre using WDM filters (not shown). An 8.2 km round trip, terminating at a WR Slave (SL), enables continuous monitoring of the stability of the WR network time using a time-interval counter (TIC). At TGV, each TN synchronises and triggers a USRP which transmits PNT radio signals, allowing the stand-alone receiver (Rx) to determine its position. During some runs, the Rx was synchronised to TN-6 for verification purposes. Copyrighted pictures reproduced from refs.[38,39,40] with permission from Orolia (WR-GM, WR-SL[38]), National Instruments (USRPs[39]), and OPNT (TNs and WR-BCs[40]).

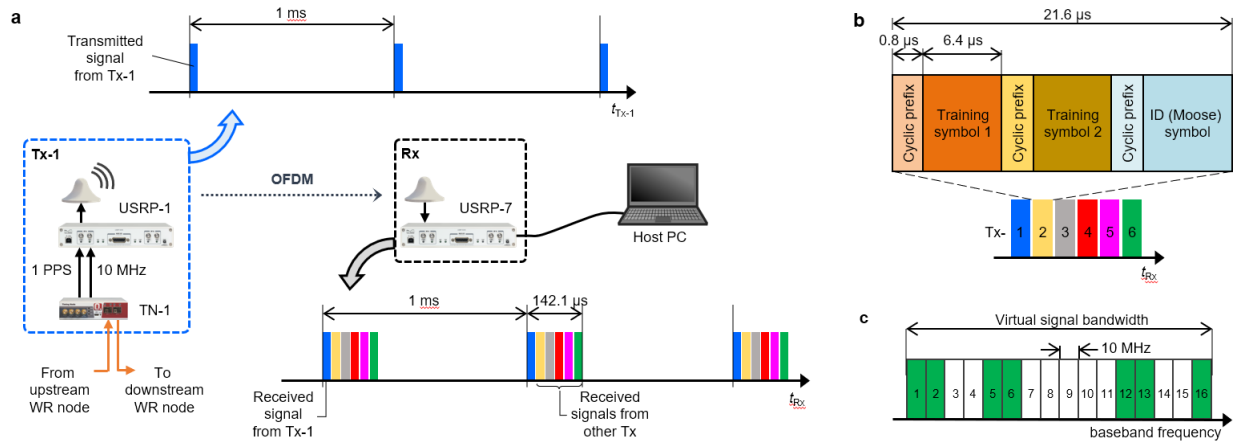

**Fig. 2| PNT radio signal structure. a**, Each TN synchronises a USRP via 10-MHz and 1-PPS signals (Tx-1 shown for example). The USRP transmits PNT signals at a 1-kHz rate; PNT signals from the six transmitters are time-division multiplexed (TDM) with 2.5-µs guard intervals. The receiver offloads the received data to a PC for offline processing. Copyrighted pictures reproduced from refs.[39,40] with permission from National Instruments (USRPs[39]) and OPNT (TN[40]). **b**, OFDM symbols transmitted during a single burst of PNT data (Methods). **c**, Multiband-OFDM is used employing either all 16 channels (160 MHz BW), or a sparse 7-band configuration with a virtual BW of 160 MHz.

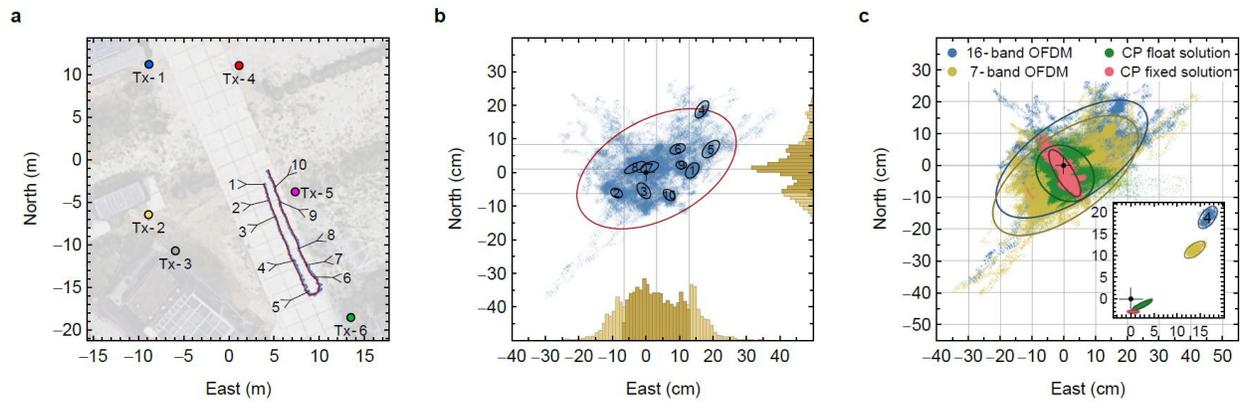

**Fig. 3|Positioning performance. a**, Aerial view (Google Earth) and trajectory of a 210-second run performed at TGV. Ten positions where the receiver was left stationary for a few seconds are indicated in chronological order. Red trace, interpolated GT trajectory; blue trace, solutions of the OFDM-TD measurements (full 160-MHz BW). **b**, Position errors (TD minus GT) for all 210,000 measurements acquired during the run (blue points). Black ellipses indicate 95% contours for 1-s subsets of the data acquired at the ten stationary points; multipath-induced biases relative to GT (black cross) are visible. The means (standard deviations) are 3.2(9.7) cm and 1.1(7.3) cm in the EW and NS directions, respectively (grey lines and yellow histograms). Red ellipse, 95th percentile of all coordinate pairs. **c**, Comparison of position errors for the various positioning methods tested here. Inset: comparison of position errors in stationary point number 4 (95% contours are shown).

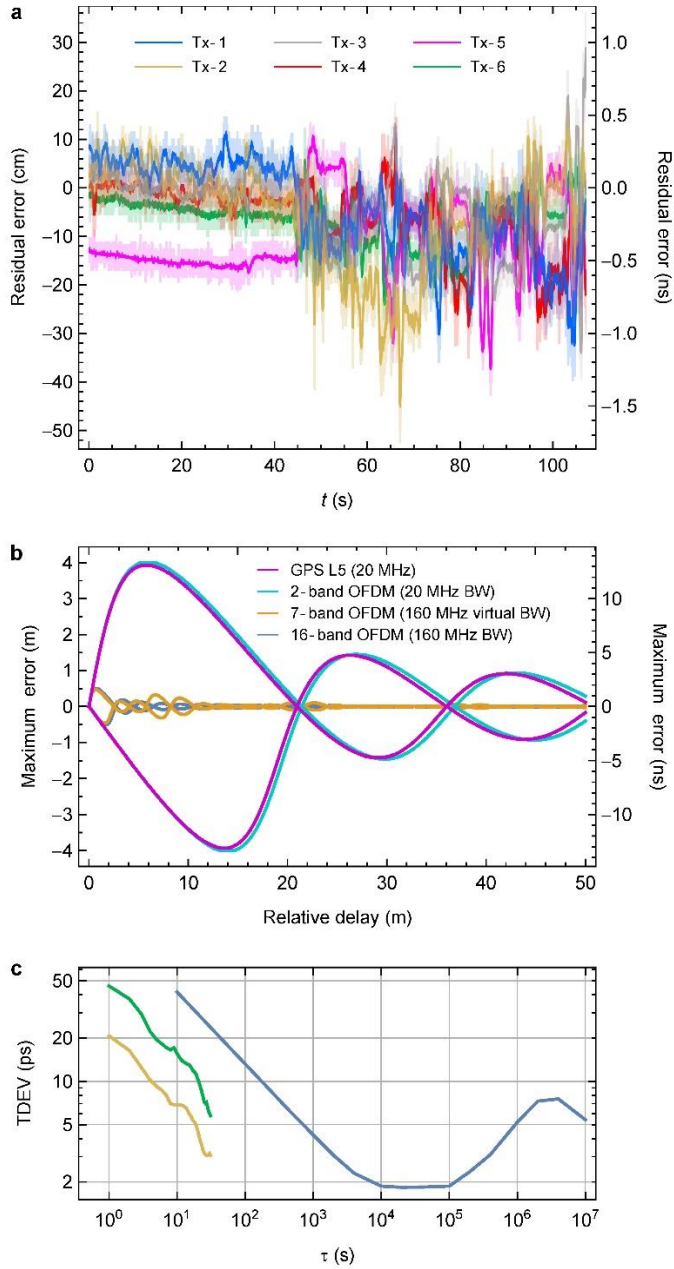

**Fig. 4|Multipath and timing performance. a**, Residual range errors measured during a test run at TGV (OFDM-TD,160-MHz BW). Lighter shades indicate unfiltered data, darker traces are moving averages obtained with a 0.3-s window. Time-varying effects of multipath occur after 50 s when the receiver is set into motion. **b**, Multipath error envelopes (MEE) for GPS L5 and various multiband OFDM signals, for TD ranging and a relative gain of the second path of 0.6. **c**, Time deviation (TDEV[26]) between the 1-PPS outputs of TN-2 and TN-1, measured at TGV (yellow). Green curve,

TDEV between TN-1 and TN-6, (daisy chain of six TNs). Blue curve, 115-day TDEV between WR-GM and WR-SL (Fig. 1**b**).

**Table 1|Positioning and timing uncertainty**

| Positioning method | RMSE (cm) | |
|---|---|---|
| | East | North |
| Time-delay positioning (full band) | 10.2 | 7.4 |
| Time-delay positioning (sparse band) | 9.4 | 8.4 |
| Carrier-phase positioning (float, full band) | 3.7 | 4.5 |
| Carrier-phase positioning (fixed, full band) | 2.2 | 4.1 |
| **Timing method** | **Standard uncertainty (ns)** | |
| WR TN time synchronisation at TGV (fibre-optic) | <0.5 | |
| Rx clock relative to WR time base (wireless) | 0.6 | |
| Rx clock relative to UTC(VSL) (wireless/fibre-optic) | 50 | |

Positioning RMSEs correspond to the data shown in Fig. 3c, and include the 2.5-cm RMSE of the ground-truth system. Residual time offsets of the WR TNs were calibrated to within 0.1 ns in the laboratory prior to installation, and verified on-site with 0.5 ns uncertainty after installation at TGV.

**Methods**

**TNPS testbed**

Extended Data Fig. 1. shows a partial map of the city of Delft, The Netherlands, indicating the locations of TGV, the various WR nodes and associated network topology, and the reference atomic clocks that are used to realize UTC(VSL) and synchronise the WR network. The network consisted of a mix of WR equipment obtained from the two WR providers mentioned in Ref. 14. The geometry of the Tx stations at TGV (Fig. 1a) was chosen for practical reasons, such as available space and suitable antenna mounting locations. This resulted into Tx-to-Tx baselines of up to 37 m, which in theory can be expanded by one order of magnitude without loss of ranging accuracy (assuming LoS conditions; see 'USRP configuration and OFDM signals' below). The setup at TGV is described in more detail by Diouf et al., who presented a preliminary analysis of positioning based on pseudo-random noise sequences instead of OFDM signals, achieving an uncertainty at the several-decimetre level[36]. At the TGV site, severe multipath conditions are expected in several locations, with reflections caused by (among other objects) a 5-m-tall billboard with a shortest distance of about 3 meters to the test trajectories, and buildings that are located 2-3 m away from Tx-1 and Tx-6. Strong reflections are also expected for the antenna of Tx-3, which is affixed to the corner of an adjacent building using a 0.2-m spacer.

Five test runs were conducted with the Rx unit carried by a trolley, and two runs were performed with the Rx unit carried in a car (with the antenna mounted on top of the roof of the car; see Extended Data Fig. 2).

**USRP configuration and OFDM signals**

PNT burst transmissions are driven at a rate of 1 kHz and a duty cycle of 0.14, with each USRP delivering 8.9 dBm to the antenna during a burst. The Rx gain is kept at 20 dB; higher gains are possible but these lead (in our setup) to occasional impairment of the received signals[36]. All USRPs use omnidirectional, ceiling mountable wideband antennas with a gain of 3 dBi. The Tx and Rx antennas are connected to the USRPs through cables of 5 m and 1 m length and cable losses of 5.6 dB and 0.9 dB, respectively. The noise level within the 160 MHz USRP BW, and referred to the input, amounts to –80.5 dBm. For a carrier frequency of 3.96 GHz, the received LoS power during a burst, $P_{Rx,LoS}$, at a distance $d$ (in metres) can be written as $P_{Rx,LoS}$ = –30.4 dBm – 20 $\log_{10}$ ($d$) dB. This translates to a coverage distance of $d$ = 320 if a signal-to-noise ratio (SNR) of 0 dB is to be achieved, which supports the estimation of pseudo-ranges with a precision of about 10 cm (CRLB).

The TDM transmission scheme, shown in Fig. 2a, was chosen for its ease of implementation in the testbed and for diagnostic purposes. The two OFDM training symbols in Fig. 2b contain identical pseudorandom noise sequences that are binary phase-shift-keyed on all available subcarriers. Both symbols are used for channel estimation, whereas only one symbol is used for packet synchronisation and ranging. The third symbol contains a Gold sequence, unique to each transmitter, for identification by the receiver. In principle, ranging and identification could be achieved using only the third symbol. When transmitting over 7 out of 16 bands, the overhead of the TDM scheme is 6.3%, and 2.5% if only the third symbol is used. A further reduction of the overhead to 0.42% would be enabled by joint transmission and detection of the PNT symbols using CDMA (analogous to GNSS signals), without significant loss of positioning performance. Alternatively, one could maintain the original PNT signals, but lower the PNT transmission rate by a factor of 160 to achieve the same rate as in 4G positioning at an overhead of 0.09%.

The 160 MHz BW of the PNT signals is limited by the maximum effective BW of the USRPs. The choice for a 3.96 GHz carrier frequency derives from a radiocommunications license for experimental research, obtained from the Dutch Telecom Agency.

**Ground-truth system and measurements**

Two robotic land-surveying total stations were used to determine the receiver's position during test runs. To this end, two 360° prisms were mounted onto the Rx vehicle, which were tracked by the total stations (see Extended Data Fig. 2). The total stations measure horizontal and vertical angles as well as slant distance, with a resulting position uncertainty of several millimetres; the GT system was also used to determine the Tx-antenna locations. Although vertical distances are taken into account in all distance measurements, the vertical coordinates of the total stations, 360° prims, and Tx and Rx antennas were determined separately and assumed not to vary during the runs on the horizontal test road. To compare trajectories obtained with the TNPS (1000 position fixes per second) and GT system (1-10 fixes per second), the GT positions were interpolated. Using the precisely known (and fixed) distance between the two 360° prisms as a length standard, the uncertainty (here expressed in terms of the RMSE) of the GT measurement was determined in one of the test runs to be about 1.8 cm; additional uncertainties due to the interpolation and small deviations of the road from the horizontal plane are estimated to increase the overall RMSE of the GT measurements to 2.5 cm (as indicated by the black cross hairs in Figs. 3b and 3c). The measured GT trajectories were time-aligned with the TNPS trajectories by offsetting the time coordinate of the GT data, such that the GT trajectory coincides in time with the trajectory found from CP measurements (note that the CP and TD trajectories are automatically time-aligned since they are derived from the same set of raw data acquired by the Rx USRP). The uncertainty introduced by this time-alignment procedure affects the obtained RMSEs only at the sub-cm level.

**Ranging and positioning through time-delay measurements**

After reception of the OFDM signals, packet synchronisation, Fourier transformation, and channel estimation, the receiver determines for each transmitter the sampled channel frequency response, *H*, of all used subcarriers[22,42]. To reduce computational complexity we employ a simplified time-delay model that considers only the LoS path, of which the delay is estimated by maximizing the cost function[22,23]

$$\hat{\tau} = \arg\max_{\tau} \frac{\boldsymbol{H}^*\boldsymbol{a}(\tau)\boldsymbol{a}^*(\tau)\boldsymbol{H}}{\boldsymbol{a}^*(\tau)\boldsymbol{a}(\tau)}. \qquad (2)$$

In Eq. (2), asterisks denote the Hermitian adjoint, and $\boldsymbol{a}$ is a complex-valued vector that describes the accumulated phase factors $\exp(-j2\pi f_p \tau)$ for each subcarrier $p$ ($p=1\ldots M\times N$).

After all six pseudo-ranges $\rho_r^i(t)$ have been determined (for a given time $t$) using Eq. (2), the receiver position $\mathbf{x}_r(t)$ and clock offset $\delta_r(t)$ are found by maximum likelihood estimation (MLE)[37], which effectively constitutes a time difference of arrival method. This can in principle be done for the three unknown receiver coordinates. However, similar to many GNSS navigation applications we are primarily interested in the horizontal position, because the test track is horizontal and the vertical coordinate (height) of the receiver antenna, $z_r$, is a known constant ($z_r$=1.573 m when placed on the trolley, and $z_r$=2.006 m when mounted onto the car). In addition, all Tx antennas are mounted at approximately the same height and at relatively small elevation angles as seen by the receiver, which makes our testbed less suitable for positioning in the vertical direction. We therefore solve only for the unknown horizontal coordinates ($x_r, y_r$) and clock offset using MLE, after linearizing the propagation distance between Tx-$i$ and Rx, $d_r^i = [(x_r - x^i)^2 + (y_r - y^i)^2 + (z_r - z^i)^2]^{1/2}$, at the receiver's approximate position.

The OFDM signals, transmitted by the six Tx-$i$ following a TDM scheme (Fig. 2a), are acquired sequentially by the receiver during a 160-μs interval, allowing sufficient time for all PNT signals to arrive. At this time scale, the free-running oscillator of the Rx USRP (a temperature-compensated crystal oscillator) is observed to drift at rates of 100-150 ps/ms. Therefore, over the course of the 160-μs acquisition period, the clock offset may drift through an interval [−12 ps, 12 ps] relative to the time-averaged clock offset. Because the same (fixed) clock offset is assumed for the determination of all six pseudo-ranges, this leads to possible errors in the pseudo-ranges in the interval [−3.6 mm, 3.6 mm]. However, these errors are negligibly small in view of other sources of ranging uncertainty.

**Hardware delays and timing uncertainty of PNT radio signals**

The variance of the pseudo-ranges depends inherently on the received SNR level and the spectral configuration of the frequency bands used. As can be seen from Eq. (1), other hardware-dependent sources of error are the uncertainties of the measured hardware delays and possible residual drifts and jitter of the oscillators in the TNs and USRPs. Hardware delays are defined relative to the 1-PPS output signal of WR timing node 1 (TN-1). Tx-*i*-specific delays $\tau'^{i}_{h}$ were determined in the laboratory using the same TN-Tx pairs and the same network configuration as used during the field trials at TGV. Delays were measured between the USRP RF output signals of the various Tx-*i* with respect to the 1-PPS output of TN-1, and values between -0.7 and 1.6 ns were found. The $\tau'^{i}_{h}$ values were furthermore found to be static (i.e. surviving tens of power cycling events of the TNs and USRPs over the course of days); typical standard deviations of repeated measurements with intermediate power cycling were 0.1-0.2 ns, indicating that the uncertainty of the $\tau'^{i}_{h}$ values is at the level of a few tenths of a nanosecond.

**Uncertainty of Rx clock offsets and wireless time distribution**

From Eq. (1) it can be seen that the common hardware delay $\tau_h$ does not need to be known for positioning, since its value can be absorbed in the receiver clock offset. However, if the goal is to relate the Rx clock to the reference clock of the TNPS (for example to use the Rx unit as a local time reference), it will be necessary to determine the clock offset $\delta_r(t)$, and consequently $\tau_h$ needs to be known. We performed an ad hoc determination of $\tau_h$ by performing a test run while the Rx USRP was synchronised by TN-6 (Fig. 1b), thus enforcing the condition $\delta_r(t)\equiv 0$. After correcting for the known delays $\tau'^i_h$ and inserting the receiver positions known from the GT system [i.e. $\mathbf{x}_r(t) \equiv \mathbf{x}_{GT}(t)$] in Eq. (1), the MLE procedure produces an apparent clock offset that is equal to $\delta_A = \tau_h + \sum_{i=1}^{6} a^i(\mathbf{x}_{GT}(t))\, \varepsilon^i_r(t)/c$, where the $a^i$ are position- and Tx-$i$-dependent sensitivity coefficients from the MLE procedure. We take the time-average of all clock offsets $\delta_A$ determined during this run, and define $\tau_h$ to be equal to the result (17,156.4 ns with a standard deviation of 0.1 ns). However, this assumes that the time-average of the summation $\sum_{i=1}^{6} a^i(\mathbf{x}_{GT}(t))\, \varepsilon^i_r(t)/c$ is zero, which may not necessarily be the case. Assuming that the measured 0.2-m-level residual errors of Fig. 4a are representative of the error terms $\varepsilon^i_r(t)$, we estimate that the summation adds another 0.2 ns of uncertainty to $\tau_h$, which forms an indication of the accuracy with which clock offsets can determined. Alternatively, one might calibrate $\tau_h$ prior to installation to reduce the uncertainty further. Even under conditions of considerable multipath, with ranging errors of ~0.5 m regularly occurring (Fig. 4b), we estimate clock-offset determination (and thereby wireless distribution of the network reference time) to be possible with 0.6 ns of uncertainty.

**Carrier-phase ranging and positioning**

Measurements of the carrier phase (CP) enable positioning with considerably higher precision than positioning based on TD measurements. Typically, CP methods measure distances with a precision corresponding to a small fraction of the carrier wavelength. Therefore, given the 7.57-cm wavelength of the 3.96-GHz carrier used in our TNPS testbed, cm-range and even mm-range precision comes within reach. The phase of the central-signal carrier can be obtained from the complex gain as determined from the sampled channel frequency response $\mathbf{H}$ [22]. However, CP measurements are inherently ambiguous, as the receiver can determine only the fractional phase in the domain [−π,π), and the integer number of cycles or wavelengths of the propagation path remains unknown. The CP results shown in Fig. 3c are nevertheless obtained through CP-only positioning. To solve the integer ambiguity, at least two measurements taken at different epochs, with uninterrupted CP tracking in between, as well as a change of geometry are needed[22]. Because the transmitter locations in our testbed are fixed, the geometry change is accomplished entirely through the receiver's motion. In a CP ambiguity-float solution, the receiver position coordinates and clock offset are estimated together with the ambiguity parameters, but treating the latter as real-valued quantities (leaving room for nonzero fractional parts due to initial phase offsets of the Tx-$i$ and Rx units). In addition, we implemented an alternative procedure that exploits the integer nature of the ambiguity[43], resulting into what is known as an ambiguity-fixed solution. This procedure considerably strengthens the position solution, and effectively turns CP measurements into very precise TD measurements.

**Runs with Rx in synchronous and asynchronous mode**

In some of the test runs conducted with the trolley (including the run shown in Fig. 3), the Rx USRP was synchronised through TN-6, also carried by the trolley (Fig. 1b). In such cases, the signal processing and positioning algorithms nonetheless processed all data as if the USRP was in free-running, asynchronous mode, where the start of the signal bursts is detected in real-time using the Schmidl-Cox algorithm. Operating the Rx unit in synchronous mode therefore offers no advantages for the positioning algorithm other than that it eliminates the (negligible) mm-level range errors associated with USRP clock drift. We also verified empirically that the positioning performance does not noticeably depend on the mode of operation (synchronous or asynchronous) of the USRP; a comparison between the results of two runs obtained with the Rx USRP in synchronous and asynchronous mode, for TD as well as CP positioning, are presented in Extended Data Fig. 3.

**WR timing performance measurements**

The absolute time offset of WR TN-1 relative to UTC(VSL) was measured using a battery-powered, portable rubidium-based atomic clock. At VSL, the portable clock was synchronised to UTC(VSL) with a delay of 26(1) ns, and then carried to TGV (here, numbers appended within parentheses represent standard uncertainties). Using a TIC, the 1-PPS output of TN-1 was found to lead the portable clock's 1-PPS output by 2(10) ns. Given the $5\times10^{-12}$ Hz/Hz uncertainty of the portable clock's frequency and the three-hour duration of the verification, the total uncertainty is estimated to be 50 ns (coverage factor $k$=1 or 68% confidence level), and consequently TN-1 was found to lag UTC(VSL) by 24(50) ns. A TIC was used to validate the synchronicity (with 0.5 ns uncertainty) of the 1-PPS output of TN-1 and the 1-PPS outputs of the other TNs in the TNPS testbed. The round-trip stability (TDEV, MDEV) between WR-GM and WR-SL at VSL (Fig. 1b) was measured using a multichannel TIC connected to their respective 1-PPS signal outputs.

**Method references**

**Data availability**

The datasets that support this manuscript are available at https://doi.org/10.34894/GFDJI1.

**Code availability**

The code to process the datasets that support this manuscript is available under the MIT-0 License at https://doi.org/10.34894/GFDJI1.

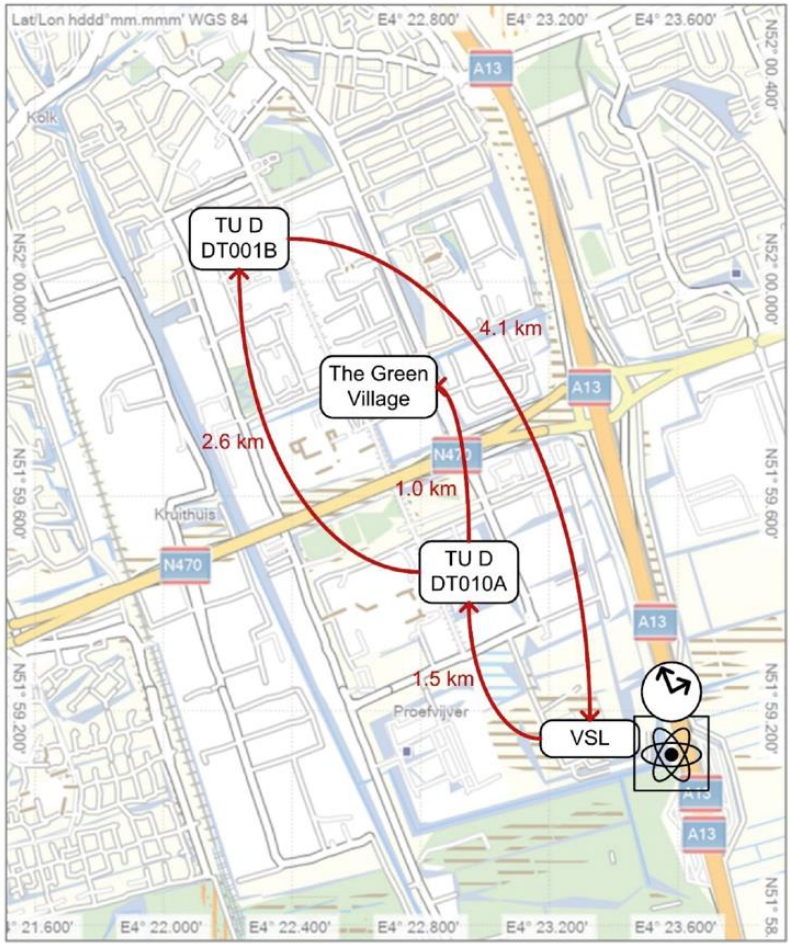

**Extended Data Fig. 1|Map of the TNPS testbed.** Locations of data centres at Delft University of Technology (TU D) and schematic representation of the fibre-optic connections are shown. The reference atomic clocks that are used to realise UTC(VSL) and synchronise the WR network are located at VSL. Map data copyright OpenStreetMap[41], obtained under Open Database License 1.0.

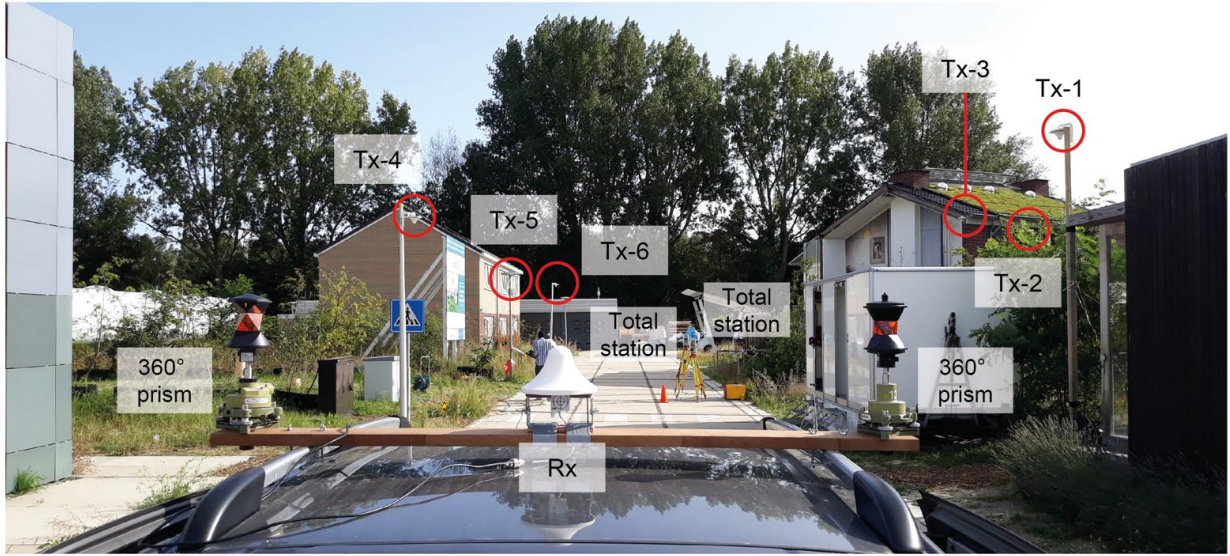

**Extended Data Fig. 2|Car setup and TGV site.** The Rx antenna and two 360° prisms are mounted onto the roof of a car. In the background part of the TGV test site is visible (viewing direction is south east). The various Tx-*i* antennas are indicated, as well as the two total stations. Tx-2 is hidden from the view by tree branches.

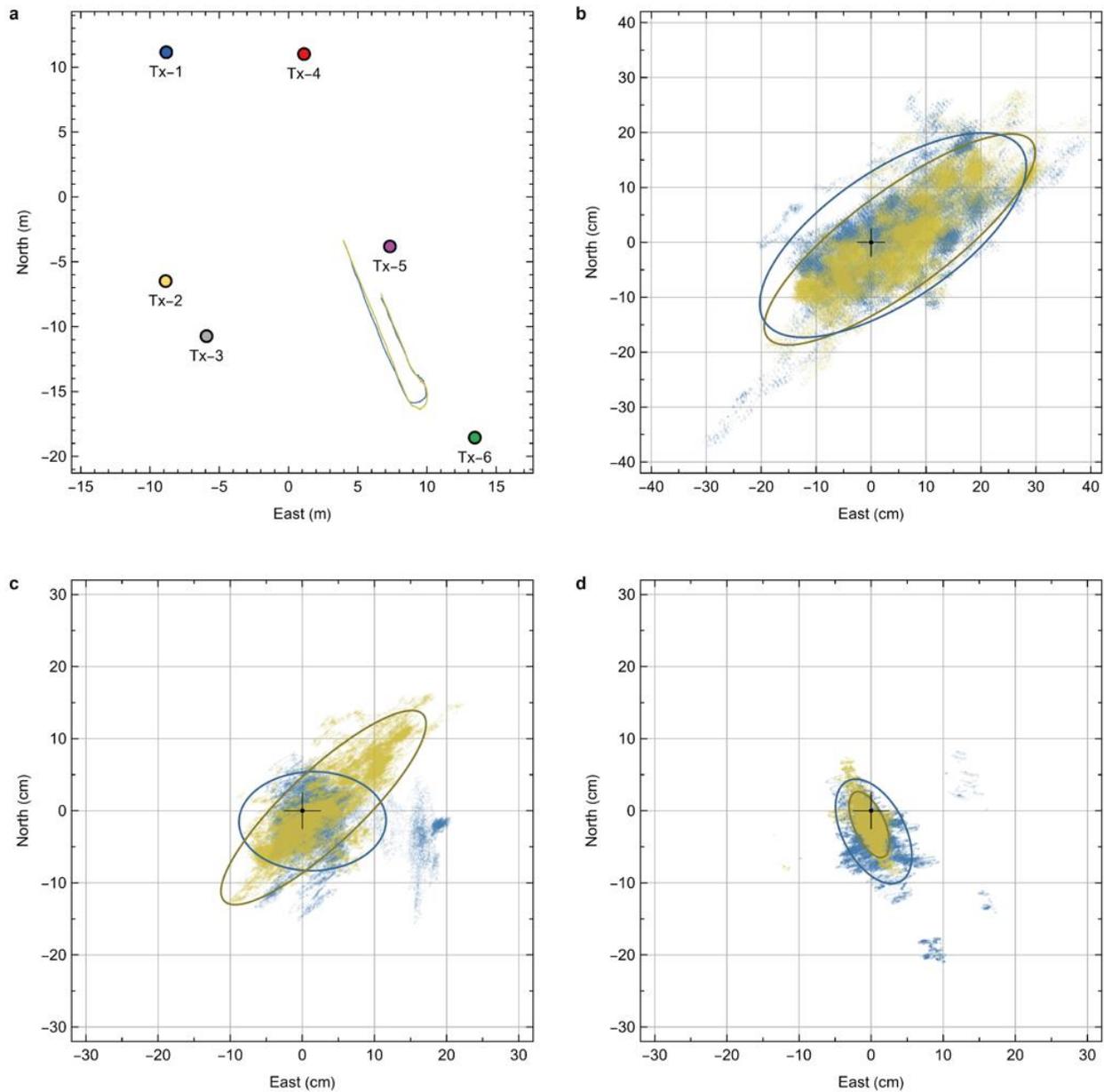

**Extended Data Fig. 3 | Comparison of positioning results in synchronous and asynchronous mode. a**, Ground-truth trajectories of two different runs, one with the Rx USRP operated in synchronous mode (blue), and one with the Rx USRP in asynchronous mode (yellow). **b**, TD position errors and 95% ellipses for the synchronous and asynchronous runs shown in **a**, following the same colour coding. The black cross indicates the GT solution and its uncertainty. **c**, Same as in **b**, but for CP ambiguity-float solutions. Note the different scale of the graph. **d**, Same as in **b**, but for

CP ambiguity-fixed solutions. During the early stages of the run in synchronous mode (blue), an incorrect integer correction was applied, leading to small islands of biased position errors.

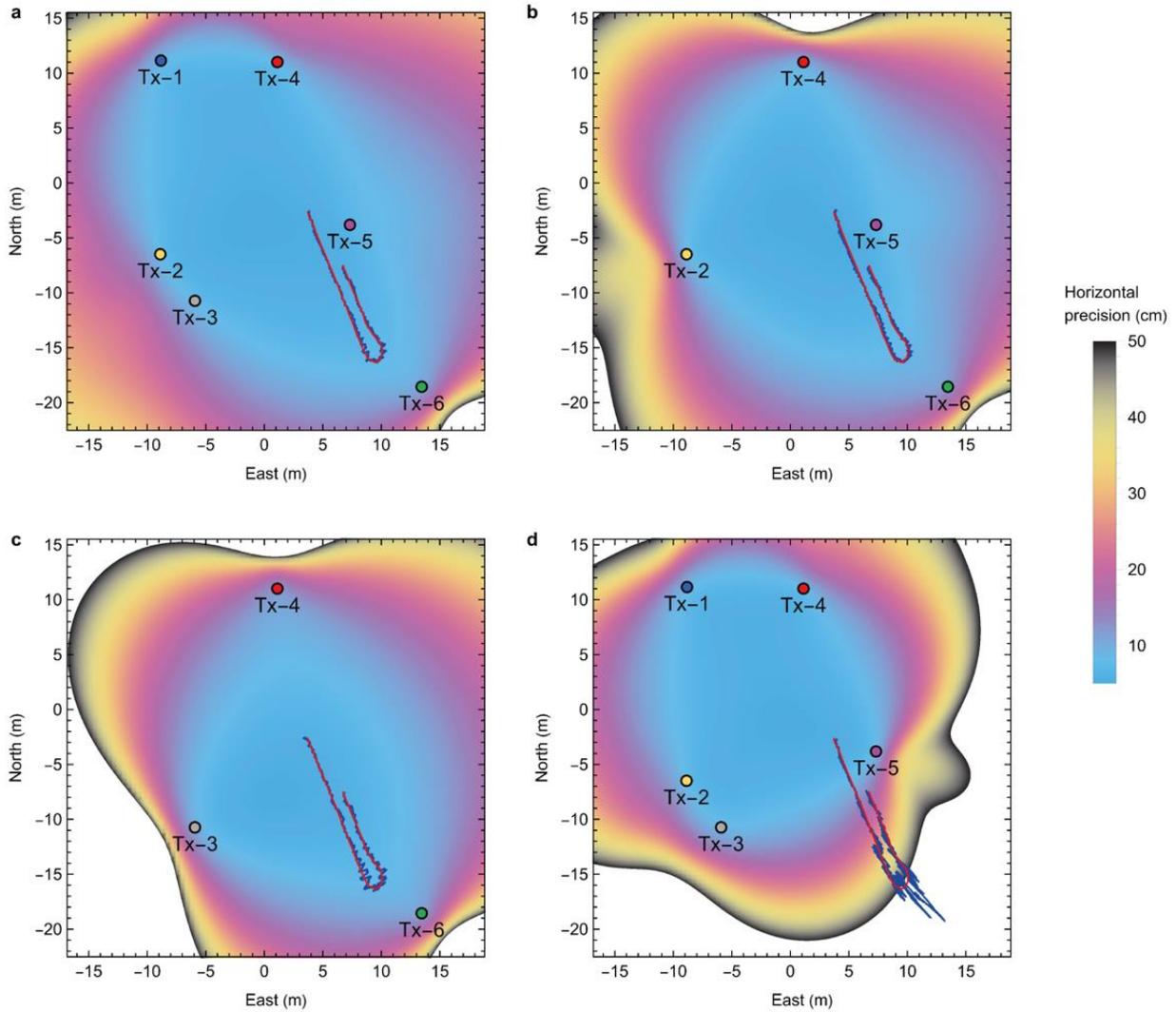

**Extended Data Fig. 4|Horizontal precision and constellation size.** Horizontal positioning precision, $(\sigma_{East}^2 + \sigma_{North}^2)^{1/2}$, with $\sigma_{East}$ and $\sigma_{North}$ the position standard deviations as determined from a nonlinear least-squares optimization that assumes normally distributed ranging errors with a standard deviation of $\sigma_r$ =6 cm for all transmitters. Values above 50 cm are clipped and replaced by white areas. Shown also are OFDM-TD position solutions for the run with the asynchronous receiver of Extended Data Fig. 3 (blue curves), and the corresponding ground-truth path (red curves). **a**, Precision and position solutions for the full TNPS constellation. **b**, Precision and position solutions for the TNPS constellation with Tx-1 and Tx-3 removed. **c**, Precision and position

solutions for the TNPS constellation with Tx-1, Tx-2, and Tx-5 removed. **d**, Precision and position solutions for the TNPS constellation with Tx-6 removed.

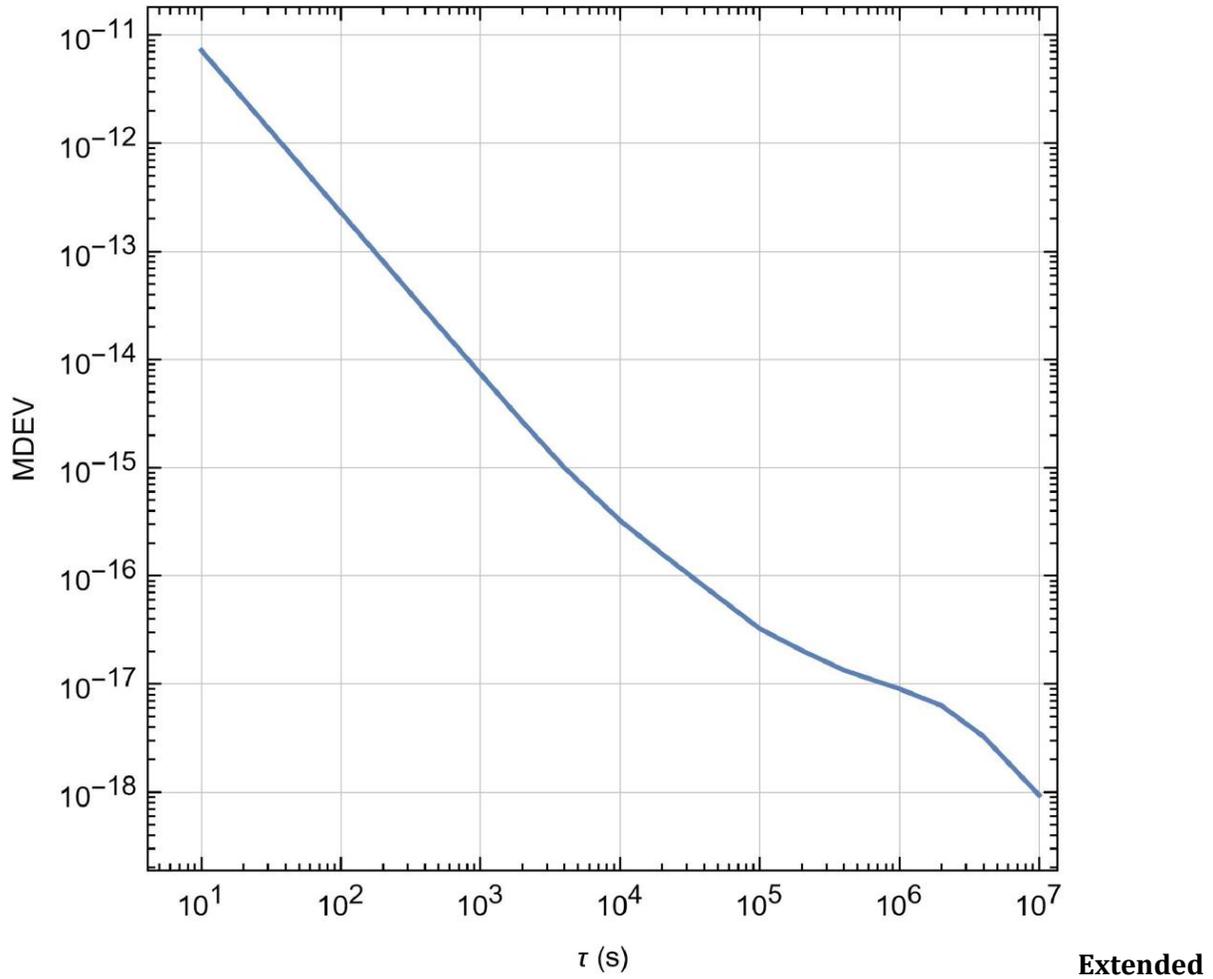

**Extended Data Fig. 5|Modified Allan deviation.** Modified Allan deviation measured between WR-GM and WR-SL at VSL (Fig. 1**b**) after a round trip through 8.2 km of installed optical fibre.

| Constellation | | | | | | Size | RMSE (cm) | | Extended Data Fig. 4 |
| Tx-1 | Tx-2 | Tx-3 | Tx-4 | Tx-5 | Tx-6 | | East | North | |
| --- | --- | --- | --- | --- | --- | --- | --- | --- | --- |
| • | • | • | • | • | • | 6 | 11.3 | 7.9 | a |
|   | • | • | • | • | • | 5 | 11.1 | 8.3 | - |
| • | • | • | • | • |   | 5 | 32.0 | 43.4 | d |
|   | • |   | • | • | • | 4 | 12.4 | 10.0 | b |
|   | • | • | • |   | • | 4 | 13.3 | 10.4 | - |
| • |   | • |   | • | • | 4 | 13.3 | 9.2 | - |
| • | • |   | • |   | • | 4 | 12.3 | 11.1 | - |
|   | • | • | • | • |   | 4 | 150 | 173 | - |
|   | • |   | • |   | • | 3 | 12.4 | 11.5 | - |
|   |   | • | • |   | • | 3 | 18.2 | 11.0 | c |
|   | • |   |   | • | • | 3 | 15.0 | 41.8 | - |
| • |   |   | • |   | • | 3 | 27.3 | 21.2 | - |

**Extended Data Table 1|TD positioning RMSE for various reduced constellations.** Values for TD positioning. Positioning RMSEs include the 2.5-cm RMSE of the ground-truth system.

| Constellation | | | | | | Size | CP float, RMSE (cm) | | CP fixed, RMSE (cm) | |
| --- | --- | --- | --- | --- | --- | --- | --- | --- | --- | --- |
| Tx-1 | Tx-2 | Tx-3 | Tx-4 | Tx-5 | Tx-6 | | East | North | East | North |
| • | • | • | • | • | • | 6 | 3.7 | 4.5 | 2.1 | 3.9 |
|  | • | • | • | • | • | 5 | 3.3 | 4.2 | 1.7 | 3.6 |
| • | • |  | • | • | • | 5 | 7.2 | 5.0 | 2.0 | 3.6 |

**Extended Data Table 2|CP positioning RMSE for various reduced constellations.** RMSEs for CP ambiguity-float and CP ambiguity-fixed solutions corresponding to the run shown in Fig. 3. Positioning RMSEs include the 2.5-cm RMSE of the ground-truth system.